\documentclass{emulateapj}
\usepackage{natbib}
\usepackage{amsmath}
\usepackage{amssymb}
\usepackage{subfigure}
\usepackage{graphicx}
\usepackage{color}

\shorttitle{The Galaxy Luminosity-Metallicity-Color Relation}
\shortauthors{Sanders et al.}

\newcommand{\photozallmed}{0.000}
\newcommand{\photozallstd}{0.008}

\newcommand{\grTzerofourmodelalphaMscatz}{0.13}

\newcommand{\grPPzerofourOthreeNtwomodelalphaMscatz}{0.10}
\newcommand{\grPPzerofourOthreeNtwomodelalphaMscatb}{0.07}
\newcommand{\grPPzerofourOthreeNtwomodelalphaMpdiff}{37}

\newcommand{\gps}{\ensuremath{g_{\rm P1}}}
\newcommand{\rps}{\ensuremath{r_{\rm P1}}}
\newcommand{\ips}{\ensuremath{i_{\rm P1}}}
\newcommand{\zps}{\ensuremath{z_{\rm P1}}}
\newcommand{\yps}{\ensuremath{y_{\rm P1}}}

\def\asec{\char'175 }

\newcommand{\kms}{{\rm km~s}^{-1}}

\begin{document}

\title{Using Colors to Improve Photometric Metallicity Estimates for Galaxies}
\author{
N. E. Sanders\altaffilmark{1}, 
E. M. Levesque\altaffilmark{2,}\altaffilmark{3},
A. M. Soderberg\altaffilmark{1}
}

\altaffiltext{1}{Harvard-Smithsonian Center for Astrophysics, 60 Garden Street, Cambridge, MA 02138 USA}
\altaffiltext{2}{CASA, Department of Astrophysical and Planetary Sciences, University of Colorado, 389-UCB, Boulder, CO 80309, USA}
\altaffiltext{3}{Einstein Fellow}

\email{nsanders@cfa.harvard.edu}

\begin{abstract}

There is a well known correlation between the mass and metallicity of star-forming galaxies.  Because mass is correlated with luminosity, this relation is often exploited, when spectroscopy is not available, to estimate galaxy metallicities based on single band photometry.  However, we show that galaxy color is typically more effective than luminosity as a predictor of metallicity.  This is a consequence of the correlation between color and the galaxy mass-to-light ratio and the recently discovered correlation between star formation rate (SFR) and residuals from the mass-metallicity relation.  
Using Sloan Digital Sky Survey spectroscopy of $\sim180,000$ nearby galaxies, we derive ``$LZC$ relations,'' empirical relations between metallicity (in seven common strong line diagnostics), luminosity, and color (in ten filter pairs and four methods of photometry).  We show that these relations allow photometric metallicity estimates, based on luminosity and a single optical color, that are $\sim50\%$ more precise 

than those made based on luminosity alone; galaxy metallicity can be estimated to within $\sim0.05-0.1$~dex of the spectroscopically-derived value depending on the diagnostic used.  Including color information in photometric metallicity estimates also reduces systematic biases for populations skewed toward high or low SFR environments, as we illustrate using the host galaxy of the supernova SN~2010ay.  This new tool will lend more statistical power to studies of galaxy populations, such as supernova and gamma-ray burst host environments, in ongoing and future wide field imaging surveys. 

\smallskip
\end{abstract}


\section{INTRODUCTION}
\label{sec:intro} 

The gas-phase metallicity of galaxies, as measured from their nebular emission spectrum, is correlated with galaxy luminosity \citep{Lequeux79,Garnett87}.  This relation has been used as a key observational tool in the study of populations such as supernova host galaxies \citep[e.g.][]{PB03,Arcavi10}, where gas-phase metallicity is an important proxy for the properties of their short lived progenitor stars.  However, using Sloan Digital Sky Survey (SDSS) imaging and spectroscopy for $\sim53,000$ galaxies, \cite{Tremonti04} showed that the luminosity-metallicity relation has a large intrinsic scatter of $\sigma=0.16$~dex (50\%), in terms of metallicity residuals, which limits the utility of this relation as an effective indicator of metallicity.

There are two primary causes for the scatter in the luminosity-metallicity relation.  First, while the scatter in the \textit{mass}-metallicity relation is fairly small ($\sigma=0.10$~dex, \citealt{Tremonti04}), luminosity is not a perfect proxy for mass.  The mass-to-light ratio of galaxies is highly correlated with galaxy color, such that redder galaxies at a fixed luminosity are more massive \citep[][]{Bell01,Kauffmann03a}.  Second, a more fundamental relation has been uncovered between mass (M), metallicity ($Z$), and star formation rate (SFR) \citep{LaraLopez10,Mannucci10}.  
This  ``fundamental plane'' or ``Fundamental Metallicity Relation'' has remarkably small residual scatter ($\sigma=0.05$~dex), indicating that variations in SFR are responsible for much of the scatter in the mass-metallicity relation.  The existence of this fundamental plane is a valuable constraint for models of galaxy evolution, and likely an expression of galaxy outflows, infall, downsizing, and/or gas-rich mergers \citep{Mannucci10,Peeples11,Yates12}.  To improve the precision of photometric metallicity estimates, a readily accessible observable must be used to break the degeneracy between luminosity, mass, and SFR.

In this paper, we show that the addition of color information significantly decreases the scatter in photometric metallicity estimates.  We derive the optimal projection of the fundamental plane for star-forming galaxies, in terms of the observable properties luminosity and color, that we call the $LZC$ relation.  We describe the sample of SDSS galaxies we use to study these correlations and methods for spectroscopic metallicity estimation in Section~\ref{sec:obs}.  In Section~\ref{sec:res}, we derive analytic expressions for the $LZC$ as expressed in a variety of different filter sets, methods of photometry, and metallicity diagnostics and we discuss the limitations of these calibrations in Section~\ref{sec:caveats}.    Finally, in Section~\ref{sec:disc}, we describe how specific observational studies may benefit from the $LZC$ relation in making precise metallicity determinations from imaging available from wide field sky surveys.

\section{GALAXY SAMPLE}
\label{sec:obs}

We used spectroscopic data and derived quantities from the MPA/JHU catalog\footnote{The MPA/JHU catalog is available at http://www.mpa-garching.mpg.de/SDSS} of $927,552$ star-forming galaxies from the SDSS-DR7 \citep{sdss7}.\footnote{We assume a standard $\Lambda$CDM cosmology throughout this work, adopting the Hubble constant $H_0=73~\kms~\rm{Mpc}^{-1}$, $\Omega_m=0.3$, and $\Lambda=0.7$.}  The catalog includes emission line fluxes, stellar masses (based on SED fitting to $ugriz$ photometry), and star formation rates for each galaxy, as described in \cite{Kauffmann03a,Brinchmann04,Salim07}.  While the MPA/JHU line fluxes are corrected for Galactic extinction, we additionally correct them for intrinsic reddening using the Balmer flux decrement. The fluxes are measured on continuum-subtracted spectra and therefore the H$\alpha$ and H$\beta$~line fluxes are corrected for Balmer absorption from the underlying stellar population \citep{Tremonti04}.  
We assume $F_{\rm{H\alpha}}/F_{\rm{H\beta}}=2.85$ (corresponding to $T=10,000$~K and $n_e=10^4~\mbox{cm}^{-3}$ for Case B recombination; \citealt{oandf}) and the extinction curve of \cite{cardelli89}, assuming $R_V=3.1$.

We joined the MPA/JHU catalog data with the photometric data from the SDSS-DR9 \citep{sdss9}.  To compare the effects of different methods of photometry, we include the SDSS model, cModel, Petrosian, and $3\asec$ fiber magnitudes \citep[for details see][]{sloanEDR}.  We adopt the model/cModel and Petrosian magnitude $K$-corrections provided in the NYU Value Added Galaxy Catalog \citep{NYUVAGC,Blanton07,Padmanabhan08}, and for the fiber magnitudes we adopt the $K$-corrections from the MPA/JHU catalog (only available for the $g,r,i$ filters).\footnote{All $K$-corrections are made to the $z=0$ frame.}  We correct the photometry for foreground Galactic dust extinction \citep{SFD}.

We perform preliminary sample cuts on the catalog following a modified version of the prescription of \cite{Mannucci10}, as follows.  First, we require that the galaxy be included in the SDSS MAIN spectroscopic sample, i.e. $r<17.77$~mag after Galactic redenning correction.  Second, we limit the sample to galaxies with $0.03<z<0.3$.  This guarantees the availability of [\ion{O}{2}]~$\lambda3727$ and is more inclusive than the $0.07<z<0.3$ cut of \citealt{Mannucci10}.  
Third,  we adopt the data quality cut from \cite{Mannucci10}, $(\rm{S/N})_{\rm{H}_\alpha}>25$, $F_{\rm{H\alpha}}/F_{\rm{H\beta}}>2.5$, which they chose to provide high data quality (high signal to noise and not saturated) in all relevant emission lines without biasing the sample explicity towards higher metallicities.  Fourth, we require the fraction of the $r$-band flux within the SDSS fiber to the full Petrosian flux to be $>0.05$, to exclude $\sim0.1\%$ galaxies where the SDSS spectroscopy includes very little of the total flux and may not reflect the galaxy global properties.  

Fifth, we reject AGN following \cite{Kauffmann03b}.  Finally, we require that $K$ corrections be available (see methodology below) and that the derived metallicitiy (see methodology below) be within a reasonable physical range ($7<\rm{log(O/H)+12}<9.5$).
We make no selection based on the internal extinction within the galaxies ($A_V$; \citealt{Mannucci10} excluded high-reddening galaxies).

For each galaxy, we compute oxygen abundance as a proxy for metallicity using a variety of strong line diagnostics that are widely used in the literature (Table~\ref{tab:diag}; see \citealt{LopezSanchez12} for a recent review).  First, we employ several diagnostics relying on the R23 ratio, which depends on the fluxes of \ion{O}{2}~$\lambda\lambda3726,3729$, \ion{O}{3}~$\lambda4959$ and $\lambda5007$, and H$\beta$.  
The R23 diagnostic suffers from a degeneracy \citep[see e.g.][]{KD02} that we break using either the N2O2 (\ion{N}{2}~$\lambda6584$ to \ion{O}{2}~$\lambda3727$) or N2 (\ion{N}{2}~$\lambda6584$ to H$\alpha$) ratios, following the authors' prescriptions.  Next we employ diagnostics depending on N2, N2O2, and O3N2, the flux ratio of \ion{O}{3}~$\lambda5007$ and \ion{N}{2}~$\lambda6584$.   
Finally, we employ diagnostics based on the ionization parameter, $P$, the ratio of \ion{O}{3}~$\lambda4959$ and $\lambda5007$ to R23.  It is necessary to calibrate for multiple diagnostics because they exhibit well known systematic discrepancies, which are particularly strong between those diagnostics calibrated empirically and those calibrated against photoionization models \citep{KE08}.  \cite{Yates12} have already shown that the fundamental metallicity relation varies with the diagnostic used.

\begin{deluxetable}{lll}
\tablecaption{Metallicity Diagnostics Used\label{tab:diag}}
\tablehead{ \colhead{Name} & \colhead{Method} & \colhead{Source}}
\scriptsize
\startdata
M91 & R23 & \citealt{M91}\tablenotemark{a}\\
KD02 & N2O2 & \citealt{KD02}\tablenotemark{b}\\
KK04 & R23 & \citealt{KK04}  \\
PP04 & N2,O3N2 & \citealt{PP04}  \\
T04 & model\tablenotemark{c} & \citealt{Tremonti04} \\
PVT & $P$ & \citealt{PVT}\tablenotemark{d}
\enddata
\tablenotetext{a}{We have adopted the revised prescription suggested by \cite{kobulnicky99}.}
\tablenotetext{b}{Following \cite{KE08}, we adopt the average of the M91 and KK04 for the lower branch solution.}
\tablenotetext{c}{The T04 metallicities are estimated based on simultaneous fits of all major emission lines to photoionization models and are provided in the MPA/JHU catalog}
\tablenotetext{d}{We use the ``ONS'' solution, which includes a dependence on the \ion{S}{2} flux, for the conditions $\log(\rm{N2})>-0.1$ and $\log(\rm{N2/S2})>-0.25$ (which is true for $\sim98\%$ of the SDSS galaxies).}

\end{deluxetable}

The number of galaxies in our final sample, following the cuts described, varies somewhat with the choice of filter set, photometric system, and metallicity diagnostic.  We consider $\sim(110-120)\times10^3$~galaxies with T04 metallicities and $\sim(160-180)\times10^3$~galaxies for other metallicity diagnostics.

\begin{figure}
\plotone{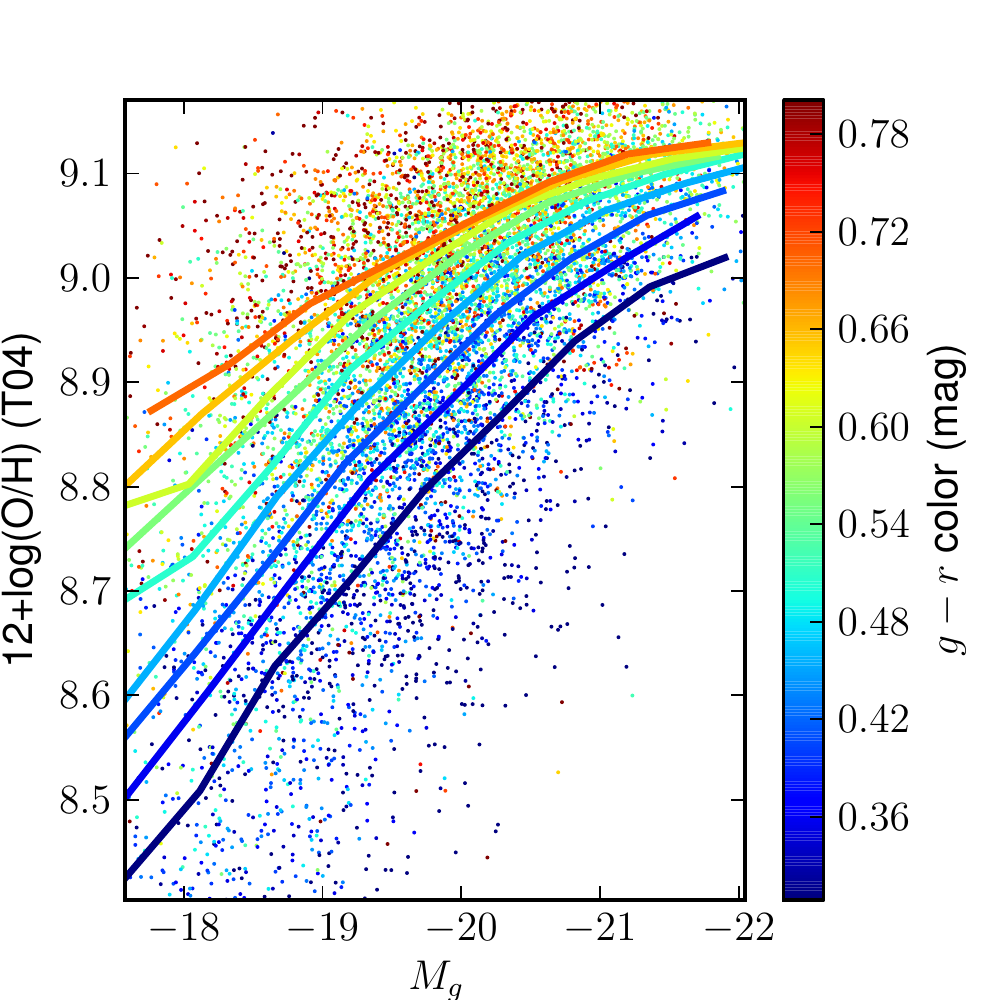}
\caption{\label{fig:Mggr}A luminosity-metallicity ($g$-band, T04 metallicity, Model magnitudes) plot showing a random subset of $10,000$ SDSS galaxies, color coded by $g-r$ color.  The solid lines show the median luminosity-metallicity relation for the galaxies divided into 10 equal-sample-size bins in $g-r$ color.}
\end{figure}

Figure~\ref{fig:Mggr} demonstrates two correlations in the galaxy sample.  First, it shows the well known luminosity-metallicity correlation (shown using $M_g$), where more luminous galaxies typically have higher metallicities.  However, there is significant scatter in this relation, with a standard deviation of {$\sigma_0=\grTzerofourmodelalphaMscatz$~dex} in the metallicity residuals from the $g$~band luminosity-metallicity relation.  Second, the figure demonstrates that there is a correlation between the residual in metallicity (the offset from the luminosity-metallicity relation) and galaxy color.

\section{LUMINOSITY-METALLICITY-COLOR RELATION}
\label{sec:res}

\begin{figure*}
\plotone{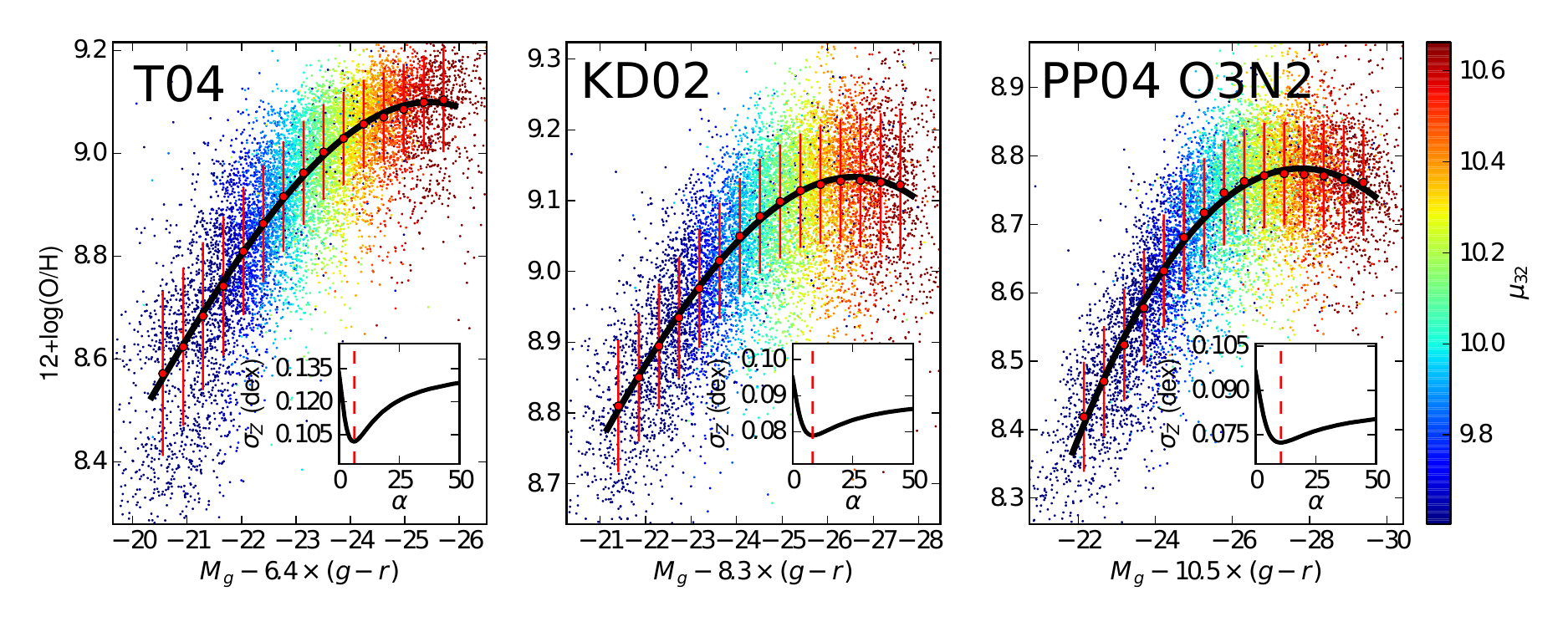}
\caption{\label{fig:Amin}The optimal projection of the $LZC$ relation for $M_g$, $g-r$ color, and three different metallicity diagnostics (T04,KD02, and PP04~O3N2; all in the Model photometric system).  The red points and lines show the median and standard deviation of the metallicity for galaxies in 15 bins.  The projected $LZC$ relation is shown for the optimal value of $\alpha$, with the best fit third order polynomial $LZC$ relation in black, extending over the calibrated range ($2-98$th percentile). The color coding shows the optical physical parameter ($\mu_{32}$) from \cite{Mannucci10}.  The insets display the standard deviation in the residuals in metallicity from the $LZC$ as a function of the color-weighting parameter $\alpha$, with the optimal value marked by the dashed line.}
\end{figure*}

Following \cite{Mannucci10}, we project the $LZC$ relation onto an axis $\mu$ with components of color and luminosity:

\begin{align}\label{eq:$LZC$}
\mu&=M_i-\alpha\times(m_i-m_j) \\
12+\log(\rm{O/H})&=p_0+p_1~\mu+p_2~\mu^2+p_3~\mu^3
\end{align}

\noindent where $i,j$ are choices of filters and $p_l$ are parameters of a third order polynomial.  For each combination of metallicity diagnostic, luminosity band, and color, we determine the optimal projection of the galaxy $LZC$ relation by sampling from a grid of $\alpha$ parameters and selecting the value that minimizes the variance in the residuals of metallicity.  We calculate the best-fit polynomial using the median value for metallicity in 15 equally-spaced bins along the projected axis, as shown in Figure~\ref{fig:Amin}.  We report the optimal value of $\alpha$ and corresponding best fit $LZC$ parameters $p$ in Table~\ref{tab:param} (for $\alpha$ values in terms of the physical parameters mass and SFR, see \citealt{Andrews13}).  

We find a correlation between the luminosity-metallicity relation and color which varies in strength depending on the choice of metallicity diagnostic and filter set.  Figure~\ref{fig:Amin} demonstrates this optimization for one filter set ($g,r$) and three choices of metallicity diagnostic: T04, KD02, and PP04~O3N2.  For PP04~O3N2, the scatter in the metallicity residuals of the $LZC$ relation is $\sigma_Z=\grPPzerofourOthreeNtwomodelalphaMscatb$~dex, as compared to the LZ relation ($\alpha=0$), $\sigma_{Z,0}=\grPPzerofourOthreeNtwomodelalphaMscatz$~dex (an improvement of $\grPPzerofourOthreeNtwomodelalphaMpdiff\%$ on a linear scale).  

The decrease in residual scatter is similar in other metallicity diagnostics, ranging from $17-51\%$. For further statistics, see Table~\ref{tab:param}.

A nominal correction for the mass-to-light ratio only will account for much, but not all, of the reduction in scatter.  For example, for $M_g$ and $(g-r)$, $\alpha=5.4$ would correspond to the mass-to-light ratio necessary to convert luminosity ($M_g$) to stellar mass \citep{Kauffmann03a}, but this value of $\alpha$ is smaller than the optimal value in any metallicity diagnostic ($\alpha\sim6-19$, see Table~\ref{tab:param}).

Note that, regardless of the choice of diagnostic or filters, the residual scatter is lower for asymptotically high values of $\alpha$ than it is for $\alpha=0$.  This implies that, in general, color is more effective than luminosity as a predictor of metallicity.

In contrast, the residual scatter achieved by \cite{Mannucci10} in terms of the optimal projection of the physical parameters stellar mass ($M_*$) and SFR, $\mu_{0.32}=\log(M_*)-0.32\log(\rm{SFR})$, was only $0.05$~dex.  However, estimation of $\mu_{0.32}$ is based on full $ugriz$ imaging and $R\sim2000$ optical spectroscopy, while the $LZC$ relation relies on imaging in just two bands and a redshift estimate.  The coloring in Figure~\ref{fig:Amin} illustrates that the optimal projection of the photometric properties is highly correlated with $\mu_{0.32}$, with Pearson correlation coefficient $\rho\sim0.6-0.9$ for all metallicity diagnostics and filters.  We calculated $\mu_{0.32}$ for the galaxies in our sample using the photometric mass measurement and aperture-corrected SFR estimates from the MPA-JHU catalog.

\begin{figure}
\plotone{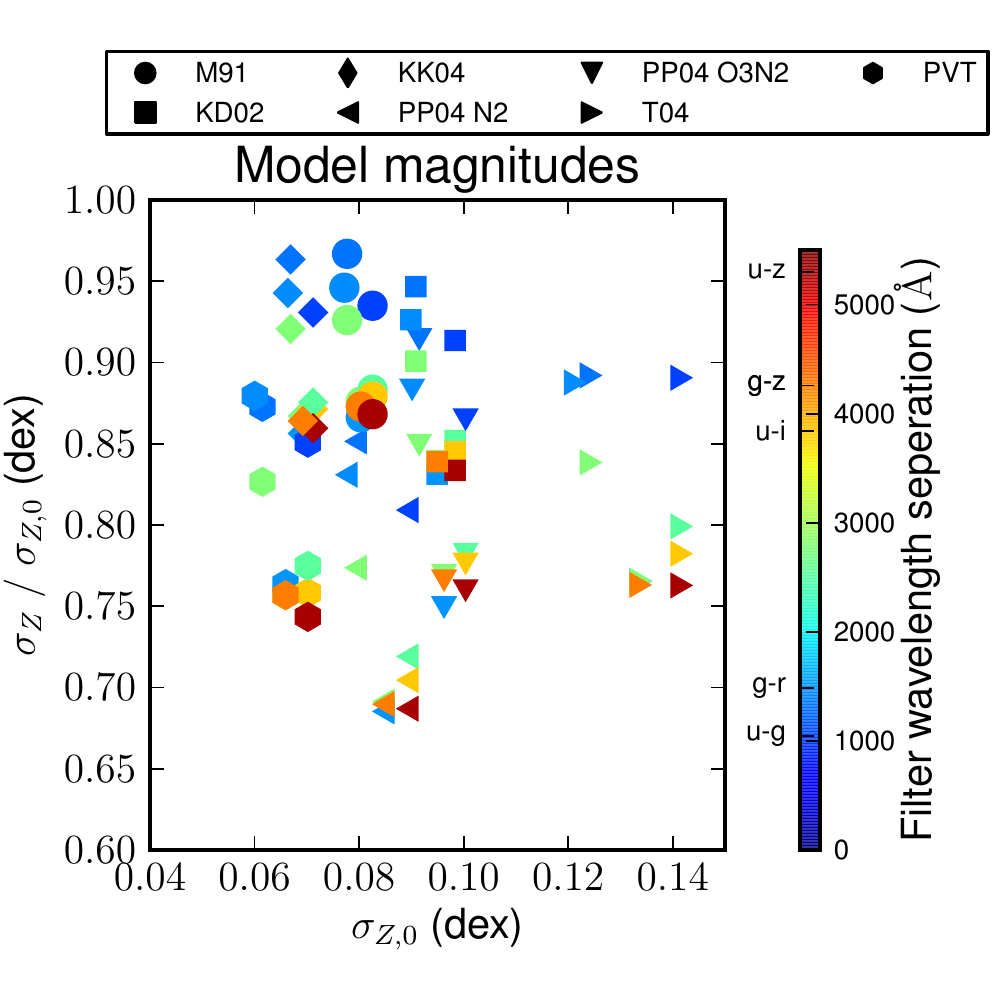}
\caption{\label{fig:ss}Summary statistics of the LZC calibrations.  The axes show $\sigma_{Z,0}$, the scatter in metallicity around the luminosity-metallicity relation, and the ratio of $\sigma_Z$, the scatter in metallicity around the optimal LZC relation, to $\sigma_{Z,0}$.  Smaller values on the $y$-axis represent improvement in scatter due to the addition of color information.  Each point shown represents an independent calibration for a particular choice of luminosity filter, color filter pair (see color coding), and metallicity diagnostic (different symbols).  The color coding is based on the seperation in Angstroms of the effective wavelength of the two color filters, and a few filter combinations are indicated in the colorbar at right.  Only results from the model photometric method is shown -- results for other methods are similar.  }
\end{figure}

In general, filter sets that include bluer filters and/or incorporate a greater separation in central wavelengths produce a greater improvement in $\sigma_Z$ (see Figure~\ref{fig:ss}).  The three most effective filter combinations are [$g-r,g-i,u-z$], producing $\sigma_Z/\sigma_{Z,0}=[0.78,0.80,0.80]$ (taking the median across all metallicity diagnostics and methods of photometry).  The three least effective are [$i-z,u-g,r-i$], with $\sigma_Z/\sigma_{Z,0}=[0.94,0.91,0.91]$.

\begin{deluxetable*}{llrrrrrrrrr}
\tablecaption{Parameters of the $LZC$ relations\label{tab:param}}
\tablehead{ \colhead{$L$} & \colhead{Color} & \colhead{$\alpha$} & \colhead{$p_3\times10^5$} & \colhead{$p_2\times10^3$} & \colhead{$p_1\times10^3$} & \colhead{$p_0$} & \colhead{$\sigma_{Z,0}$} & \colhead{$\sigma_{Z}$} & \colhead{$\sigma_{Z}/\sigma_{Z,0}$} & \colhead{$\mu$ range} }
\scriptsize
\startdata
\sidehead{PP04 O3N2 -- model} 
$u$ &$u-g$ &5.8 &-98.57 &-94.43 &-2951.55 &-21.48 &0.100 &0.087 &0.867 &[-29.0,-22.8] \\
$u$ &$u-r$ &4.1 &-55.79 &-59.71 &-2023.09 &-13.30 &0.100 &0.079 &0.784 &[-29.4,-22.7] \\
$u$ &$u-i$ &3.9 &-83.33 &-82.75 &-2687.82 &-19.90 &0.100 &0.078 &0.778 &[-30.5,-23.5] \\
$u$ &$u-z$ &3.5 &-47.07 &-51.13 &-1769.68 &-11.02 &0.100 &0.076 &0.762 &[-30.6,-23.0] \\
$g$ &$g-r$ &10.5 &-9.02 &-18.82 &-836.54 &-1.87 &0.096 &0.072 &0.752 &[-29.7,-21.9] \\
$g$ &$g-i$ &7.5 &-68.65 &-68.11 &-2210.54 &-14.79 &0.096 &0.074 &0.771 &[-30.5,-23.0] \\
$g$ &$g-z$ &6.2 &-19.35 &-25.61 &-993.07 &-3.20 &0.096 &0.074 &0.768 &[-30.9,-22.3] \\
$r$ &$r-i$ &11.0 &-241.03 &-198.54 &-5431.00 &-40.58 &0.092 &0.084 &0.916 &[-26.9,-21.7] \\
$r$ &$r-z$ &9.6 &-16.28 &-24.51 &-977.73 &-2.96 &0.092 &0.078 &0.851 &[-29.2,-21.6] \\
$i$ &$i-z$ &11.3 &84.65 &45.53 &699.29 &11.03 &0.090 &0.080 &0.885 &[-26.1,-19.1] \enddata
\tablecomments{Parameters of the $LZC$ relation defined in Equation~\ref{eq:$LZC$}; $L$ is the photometric band of the luminosity, $\alpha$ is the optimal value of the color weighting factor to minimize the scatter in metallicity, and $p_l$ are the parameters of the best-fit $LZC$ polynomial for the corresponding value of $\alpha$.  The value $\sigma_{Z}/\sigma_{Z,0}$ expresses the reduction in the scatter relative to the LZ relation (without color term).  The range of $\mu$, the optimal projection of luminosity and color, over which the diagnostic is calibrated.  The $\mu$ range is defined by the 2nd and 98th percentiles in $\mu$ (see Equation~\ref{eq:$LZC$}) of the SDSS galaxies in the calibration sample.  Table~\ref{tab:param} is published in its entirety in the electronic edition, including parameters for all metallicity diagnostics.  A portion is shown here for guidance regarding its form and content.}
\end{deluxetable*}

\section{CAVEATS}
\label{sec:caveats}

Here we note certain caveats and limitations of the $LZC$ calibrations presented in this work and caution users not to apply them outside of the regime of the calibration data.

First, we recommend that the SDSS model magnitudes be used when applying the $LZC$ relations to estimate galaxy metallicity.  Model magnitudes should provide the most accurate measurements of galaxy colors\footnote{http://www.sdss3.org/dr9/algorithms/magnitudes.php)}, although Petrosian and cModel magnitudes are typically in agreement with model magnitudes to within $<0.1$~mag.  We provide calibrations using the other SDSS photometric methods here for completeness and to support applications to datasets where photometry is only available in a particular method (i.e. Petrosian photometry).  Fiber magnitudes, which are integrated over a fixed $3\asec$ aperture, may not encompass the full galaxy for large or nearby objects.  The fiber magnitude calibrations may be useful for explorations of aperture effects \citep[see e.g.][]{Kewley05}.

Second, the redshift range of the calibration data is $0.03<z<0.3$ (Section~\ref{sec:obs}).  The $LZC$ relations need to be tested for evolution at higher redshifts due to evolution in the fundamental plane for star-forming galaxies and passive evolution of galaxy colors.  It is unclear to what extent the fundamental plane evolution would effect the calibrations.  \cite{Cresci12} found no evolution to $z\sim0.8$ and \cite{LaraLopez10} concluded that there is no detectable evolution out to $z\sim3.5$.  However, \cite{PerezM12} investigate a larger sample of $\sim5000$ zCOSMOS galaxies to $z\sim1.3$ and report evolution of the SFR-corrected mass-metallicity relation starting at $z\gtrsim0.4$.  As data is assembled to corroborate evolution in these fundamental plane relations, it may also be used to calibrate the redshift dependence of the $LZC$ relations.

Third, redshift estimates are needed to evaluate luminosity and apply $K$-corrections to estimate metallicity using the rest-frame $LZC$ calibrations we present here.  Using $ugriz$ photometry of SDSS main sample galaxies ($r<17.77$~mag, $z<0.4$), photometric redshift estimates for galaxies can be achieved with scatter $\delta z\sim0.02$ \citep{Ball08}, while $K$-corrections can be determined to $\lesssim20\%$ \citep{Blanton07}.  In some applications of the $LZC$, galaxy redshift may already be known through e.g. observations of a hosted supernova.  When using only 2 bands of photometry (the minimal use case for the $LZC$), $K$-corrections have a larger uncertainty (an additional scatter of $\sim5-20\%$ versus full-photometric corrections; \citealt{Chilingarian10}).

To test the uncertainty in metallicity introduced by use of photometric redshifts (photo-$z$) and 2-band $K$-corrections, we recompute metallicities for the subset of $\sim70,000$ galaxies in the MPA-JHU catalog with photo-$z$ estimates in the SDSS-DR9.  We use the $kd$-tree nearest neighbor fit photo-$z$ estimates, as described in \cite{Csabai07}.  
We compute $K$-corrections using the analytic prescriptions of \cite{Chilingarian10}, using both the spectroscopic and photometric redshifts for each galaxy.  We apply the $LZC$ relation as calibrated for the PP04~O3N2 metallicity diagnostic using the $M_g$ luminosities and $g-r$ colors (model magnitudes).  The resulting distribution of metallicity residuals ($\delta Z$) for the spectroscopic versus photometric redshifts suggests that there is no systematic bias introduced by photo-$z$ (median $\delta Z=\photozallmed$~dex).  The typical uncertainty added to the metallicity estimates is negligible (standard deviation $\delta Z=\photozallstd$~dex) and therefore the 
photometric metallicity estimate is dominated 
by the scatter in the $LZC$ relation.

Some additional properties of galaxies may effect their metallicity as estimated from the $LZC$ relation.  Edge-on galaxies may be redder than face-on equivalents, leading towards an upward bias in their $LZC$ metallicity.  Galaxy inclination could be included as an additional parameter in a future calibration of the $LZC$.   Early-type galaxies may contaminate photometric samples of star-forming galaxies.  Early-types should be excluded by careful application of color-magnitude diagrams; they may not lie along the extrapolation of the $LZC$ relation to redder colors, and the gas-phase metallicity of early-types may not be of interest in any case.

Fourth, we caution that the statistics presented here describe the bulk of the galaxy distribution (e.g. median and $1\sigma$ contours), while individuals may be outliers from this population.  Like any photometric method for galaxy metallicity estimation, the $LZC$ relations are most robust when applied to a statistical sample of galaxies.  

\section{APPLICATIONS}
\label{sec:disc}

The calibrated $LZC$ relation we present could benefit several disciplines, allowing for precise and accurate metallicity estimates for galaxies based on photometry alone.  Spectroscopic metallicity measurement demands considerably more observational resources, while full SED modeling provides only weak constraints on metallicity ($1\sigma$ scatter of $\sim0.2$~dex, e.g. \citealt{Pacifici12}) and accesses the chemical composition in the older, stellar population rather than the gas phase.  In contrast, the $LZC$ relation can be employed to make precise measurements of gas phase metallicity using existing multi-band photometry from wide-field surveys such as SDSS or newly acquired, targeted observations, so long as an estimate for the redshift of each galaxy is available.

Studies of supernova (SN) host galaxies can support inferences into progenitor star populations \citep[e.g.][]{Modjaz08,Prieto08,nesBS}, with some studies relying on photometry rather than spectroscopy to measure the host galaxy metallicity \citep[e.g.][]{PB03,BP09,Arcavi10}. However, \cite{nesBS} have shown that the statistical uncertainty associated with metallicity estimates based on the galaxy luminosity-metallicity relation is a significant barrier to detecting subtle differences in metallicity among SN populations.  Moreover, because some SNe strongly prefer blue galaxies (high SFR environments; \citealt{Levesque10b,Kelly11}), photometric metallicity estimates will be biased if color information is not incorporated.  Using the $LZC$ relation will effectively remove this bias, and significantly reduce the uncertainty in metallicity measurements.  
The additional uncertainty introduced by photo-$z$ should be minor, as SN host galaxies studies are almost exclusively done in the $z<0.15$ regime \citep[see 
compilation in][]{nesBS}, and spectroscopic redshift estimates are often available from the SN spectroscopy.  We note that the $LZC$ relation predicts the galaxy metallicity in the inner few kpc of the galaxy, as probed by the $3\asec$ SDSS spectroscopic fibers, and significant offsets may exist from the SN host environment metallicity due to metallicity gradients in galaxies.  However, these metallicity offsets are typically small ($\lesssim0.1$~dex; \citealt{nesBS}), and the intrinsic scatter in the radial metallicity profiles of galaxies limits observers' ability to spatially isolate the explosion site with spectroscopy \citep{Rosolowsky08,SandersM31}.

Similarly, galaxy metallicity measurements are key to the discussion of the progenitor properties of long-duration gamma-ray bursts (LGRBS, see e.g. \citealt{Fynbo03,nes2010ay}).  While the host environments of LGRBs typically fall below the mass-metallicity relation \citep{Levesque10b}, it has been shown that these host galaxies do follow the fundamental plane relation (\citealt{Mannucci11}, but see also \citealt{Kocevski11}).  Because LGRBs are frequently discovered at high redshift ($z>1$), the $LZC$ relation could potentially be used to derive metallicity estimates at considerably lower expense than deep NIR spectroscopy \citep[e.g.][]{Maiolino08}.  However, in order to study the high redshift extremely lowest metallicity environments preferred by LGRBs ($Z\lesssim0.3~Z_\odot$, see e.g. \citealt{Mannucci11}), additional calibration is needed to extend the $LZC$ relation beyond the range probed by the SDSS galaxy sample, which is $0.4~Z_\odot\lesssim Z\lesssim1.3~Z_\odot$ and $z<0.3$ (with only $10\%$ of the galaxies in our sample being at $0.2<z<0.3$).

As a usage example, we apply the $LZC$ relation to the unusual host galaxy of the SN~2010ay.  In \cite{nes2010ay}, we report that this host galaxy is a $2\sigma$ outlier from the luminosity-metallicity relation.  The median and $1\sigma$ metallicity interval for SDSS galaxies with luminosity similar to this host galaxy ($M_B=-18.3$~mag) is $12+\log(\rm{O/H})=8.93\pm0.17$ (T04).  This is a factor of $\gtrsim2$ greater than the spectroscopically-measured T04 metallicity of $12+\log(\rm{O/H})=8.58$.  The discrepancy is due to the extremely low mass-to-light ratio and high SFR of the host galaxy. 
The $LZC$ relation cannot be applied in all filter combinations because the host lies outside the calibrated range for $\mu$.
Using the $LZC$ relation for $M_r$ and $r-i$ color, we find a T04 metallicity of $12+\log(\rm{O/H})=8.58\pm0.03$ (with an additional systematic uncertainty of 0.1~dex from the spread in the $LZC$ relation).  This agrees well with the spectroscopically measured value and has a significantly lower associated uncertainty than the estimate from the luminosity-metallicity relation.  
Because the LZC relation cannot be applied in all filter combinations, this cases illustrates the importance of extending the calibration presented here to lower-metallicity host galaxies not well-represented in the SDSS spectroscopic sample.

Finally, we suggest that the next generation of wide field, multi-band, photometric surveys could use the $LZC$ relation to characterize the metallicity distribution of galaxies in the local universe, and perhaps its evolution with redshift.  The Pan-STARRS1 (PS1) survey is already operating, and will provide $\gps~\rps~\ips~\zps~\yps$ photometry for $\sim2\times10^8$ galaxies over $3/4$ of the sky \citep{Saglia12}.  In the future, the Large Synoptic Survey Telescope (LSST) will provide $ugrizy$ photometry of $\sim10^{10}$ galaxies to $z\sim6$ \citep{LSST}.  With the advent of such datasets, the $LZC$ relation may play an important role in defining the metallicity distribution of galaxies that has emerged from the cosmic evolution of star formation and galaxy mass.  To fulfill that role, the calibrations presented here must first be extended to higher redshift using data from ongoing spectroscopic surveys of the high redshift universe.

\acknowledgements
\label{sec:ackn}

We thank an anonymous referee for very helpful suggestions.  We thank the MPA/JHU teams for making available their catalog of measured properties for SDSS galaxies, and we thank E. Berger, D. Eisenstein, D. Erb, R. Foley, and A. Tripathi for helpful conversations.  This work was supported by the National Science Foundation through a Graduate Research Fellowship provided to N.E.S.; E.M.L. is supported by NASA through Einstein Postdoctoral Fellowship grant number PF0-110075 awarded by the Chandra X-ray Center, which is operated by the Smithsonian Astrophysical Observatory for NASA under contract NAS8-03060; and support for this work was provided by the David and Lucile Packard Foundation Fellowship for Science and Engineering awarded to A.M.S.

{\it Facilities:} \facility{PS1}, \facility{EVLA}, \facility{Swift}, \facility{MMT}

\bibliographystyle{fapj}

\begin{thebibliography}{54}
\expandafter\ifx\csname natexlab\endcsname\relax\def\natexlab#1{#1}\fi

\bibitem[{{Abazajian} {et~al.}(2009){Abazajian}, {Adelman-McCarthy},
  {Ag{\"u}eros}, {Allam}, {Allende Prieto}, {An}, {Anderson}, {Anderson},
  {Annis}, {Bahcall}, \& et~al.}]{sdss7}
{Abazajian}, K.~N., {et~al.} 2009, \apjs, 182, 543

\bibitem[{{Ahn} {et~al.}(2012){Ahn}, {Alexandroff}, {Allende Prieto},
  {Anderson}, {Anderton}, {Andrews}, {Aubourg}, {Bailey}, {Balbinot}, {Barnes},
  \& et~al.}]{sdss9}
{Ahn}, C.~P., {et~al.} 2012, \apjs, 203, 21

\bibitem[{{Andrews} \& {Martini}(2013)}]{Andrews13}
{Andrews}, B.~H., \& {Martini}, P. 2013, \apj, 765, 140

\bibitem[{{Arcavi} {et~al.}(2010){Arcavi}, {Gal-Yam}, {Kasliwal}, {Quimby},
  {Ofek}, {Kulkarni}, {Nugent}, {Cenko}, {Bloom}, {Sullivan}, {Howell},
  {Poznanski}, {Filippenko}, {Law}, {Hook}, {J{\"o}nsson}, {Blake}, {Cooke},
  {Dekany}, {Rahmer}, {Hale}, {Smith}, {Zolkower}, {Velur}, {Walters},
  {Henning}, {Bui}, {McKenna}, \& {Jacobsen}}]{Arcavi10}
{Arcavi}, I., {et~al.} 2010, \apj, 721, 777

\bibitem[{{Ball} {et~al.}(2008){Ball}, {Brunner}, {Myers}, {Strand}, {Alberts},
  \& {Tcheng}}]{Ball08}
{Ball}, N.~M., {Brunner}, R.~J., {Myers}, A.~D., {Strand}, N.~E., {Alberts},
  S.~L., \& {Tcheng}, D. 2008, \apj, 683, 12

\bibitem[{{Bell} \& {de Jong}(2001)}]{Bell01}
{Bell}, E.~F., \& {de Jong}, R.~S. 2001, \apj, 550, 212

\bibitem[{{Blanton} \& {Roweis}(2007)}]{Blanton07}
{Blanton}, M.~R., \& {Roweis}, S. 2007, \aj, 133, 734

\bibitem[{{Blanton} {et~al.}(2005){Blanton}, {Schlegel}, {Strauss},
  {Brinkmann}, {Finkbeiner}, {Fukugita}, {Gunn}, {Hogg}, {Ivezi{\'c}}, {Knapp},
  {Lupton}, {Munn}, {Schneider}, {Tegmark}, \& {Zehavi}}]{NYUVAGC}
{Blanton}, M.~R., {et~al.} 2005, \aj, 129, 2562

\bibitem[{{Boissier} \& {Prantzos}(2009)}]{BP09}
{Boissier}, S., \& {Prantzos}, N. 2009, \aap, 503, 137

\bibitem[{{Brinchmann} {et~al.}(2004){Brinchmann}, {Charlot}, {White},
  {Tremonti}, {Kauffmann}, {Heckman}, \& {Brinkmann}}]{Brinchmann04}
{Brinchmann}, J., {Charlot}, S., {White}, S.~D.~M., {Tremonti}, C.,
  {Kauffmann}, G., {Heckman}, T., \& {Brinkmann}, J. 2004, \mnras, 351, 1151

\bibitem[{{Cardelli} {et~al.}(1989){Cardelli}, {Clayton}, \&
  {Mathis}}]{cardelli89}
{Cardelli}, J.~A., {Clayton}, G.~C., \& {Mathis}, J.~S. 1989, \apj, 345, 245

\bibitem[{{Chilingarian} {et~al.}(2010){Chilingarian}, {Melchior}, \&
  {Zolotukhin}}]{Chilingarian10}
{Chilingarian}, I.~V., {Melchior}, A.-L., \& {Zolotukhin}, I.~Y. 2010, \mnras,
  405, 1409

\bibitem[{{Cresci} {et~al.}(2012){Cresci}, {Mannucci}, {Sommariva}, {Maiolino},
  {Marconi}, \& {Brusa}}]{Cresci12}
{Cresci}, G., {Mannucci}, F., {Sommariva}, V., {Maiolino}, R., {Marconi}, A.,
  \& {Brusa}, M. 2012, \mnras, 421, 262

\bibitem[{{Csabai} {et~al.}(2007){Csabai}, {Dobos}, {Trencs{\'e}ni},
  {Herczegh}, {J{\'o}zsa}, {Purger}, {Budav{\'a}ri}, \& {Szalay}}]{Csabai07}
{Csabai}, I., {Dobos}, L., {Trencs{\'e}ni}, M., {Herczegh}, G., {J{\'o}zsa},
  P., {Purger}, N., {Budav{\'a}ri}, T., \& {Szalay}, A.~S. 2007, Astronomische
  Nachrichten, 328, 852

\bibitem[{{Fynbo} {et~al.}(2003){Fynbo}, {Jakobsson}, {M{\"o}ller}, {Hjorth},
  {Thomsen}, {Andersen}, {Fruchter}, {Gorosabel}, {Holland}, {Ledoux},
  {Pedersen}, {Rhoads}, {Weidinger}, \& {Wijers}}]{Fynbo03}
{Fynbo}, J.~P.~U., {et~al.} 2003, \aap, 406, L63

\bibitem[{{Garnett} \& {Shields}(1987)}]{Garnett87}
{Garnett}, D.~R., \& {Shields}, G.~A. 1987, \apj, 317, 82

\bibitem[{{Kauffmann} {et~al.}(2003{\natexlab{a}}){Kauffmann}, {Heckman},
  {Tremonti}, {Brinchmann}, {Charlot}, {White}, {Ridgway}, {Brinkmann},
  {Fukugita}, {Hall}, {Ivezi{\'c}}, {Richards}, \& {Schneider}}]{Kauffmann03b}
{Kauffmann}, G., {et~al.} 2003{\natexlab{a}}, \mnras, 346, 1055

\bibitem[{{Kauffmann} {et~al.}(2003{\natexlab{b}}){Kauffmann}, {Heckman},
  {White}, {Charlot}, {Tremonti}, {Brinchmann}, {Bruzual}, {Peng}, {Seibert},
  {Bernardi}, {Blanton}, {Brinkmann}, {Castander}, {Cs{\'a}bai}, {Fukugita},
  {Ivezic}, {Munn}, {Nichol}, {Padmanabhan}, {Thakar}, {Weinberg}, \&
  {York}}]{Kauffmann03a}
------. 2003{\natexlab{b}}, \mnras, 341, 33

\bibitem[{{Kelly} \& {Kirshner}(2011)}]{Kelly11}
{Kelly}, P.~L., \& {Kirshner}, R.~P. 2011, ArXiv e-prints, 1110.1377

\bibitem[{{Kewley} \& {Dopita}(2002)}]{KD02}
{Kewley}, L.~J., \& {Dopita}, M.~A. 2002, \apjs, 142, 35

\bibitem[{{Kewley} \& {Ellison}(2008)}]{KE08}
{Kewley}, L.~J., \& {Ellison}, S.~L. 2008, \apj, 681, 1183

\bibitem[{{Kewley} {et~al.}(2005){Kewley}, {Jansen}, \& {Geller}}]{Kewley05}
{Kewley}, L.~J., {Jansen}, R.~A., \& {Geller}, M.~J. 2005, \pasp, 117, 227

\bibitem[{{Kobulnicky} {et~al.}(1999){Kobulnicky}, {Kennicutt}, \&
  {Pizagno}}]{kobulnicky99}
{Kobulnicky}, H.~A., {Kennicutt}, Jr., R.~C., \& {Pizagno}, J.~L. 1999, \apj,
  514, 544

\bibitem[{{Kobulnicky} \& {Kewley}(2004)}]{KK04}
{Kobulnicky}, H.~A., \& {Kewley}, L.~J. 2004, \apj, 617, 240

\bibitem[{{Kocevski} \& {West}(2011)}]{Kocevski11}
{Kocevski}, D., \& {West}, A.~A. 2011, \apjl, 735, L8+

\bibitem[{{Lara-L{\'o}pez} {et~al.}(2010){Lara-L{\'o}pez}, {Cepa},
  {Bongiovanni}, {P{\'e}rez Garc{\'{\i}}a}, {Ederoclite}, {Casta{\~n}eda},
  {Fern{\'a}ndez Lorenzo}, {Povi{\'c}}, \& {S{\'a}nchez-Portal}}]{LaraLopez10}
{Lara-L{\'o}pez}, M.~A., {et~al.} 2010, \aap, 521, L53+

\bibitem[{{Lequeux} {et~al.}(1979){Lequeux}, {Peimbert}, {Rayo}, {Serrano}, \&
  {Torres-Peimbert}}]{Lequeux79}
{Lequeux}, J., {Peimbert}, M., {Rayo}, J.~F., {Serrano}, A., \&
  {Torres-Peimbert}, S. 1979, \aap, 80, 155

\bibitem[{{Levesque} {et~al.}(2010){Levesque}, {Kewley}, {Berger}, \&
  {Zahid}}]{Levesque10b}
{Levesque}, E.~M., {Kewley}, L.~J., {Berger}, E., \& {Zahid}, H.~J. 2010, \aj,
  140, 1557

\bibitem[{{Lopez-Sanchez} {et~al.}(2012){Lopez-Sanchez}, {Dopita}, {Kewley},
  {Zahid}, {Nicholls}, \& {Scharwachter}}]{LopezSanchez12}
{Lopez-Sanchez}, A.~R., {Dopita}, M.~A., {Kewley}, L.~J., {Zahid}, H.~J.,
  {Nicholls}, D.~C., \& {Scharwachter}, J. 2012, ArXiv e-prints, 1203.5021

\bibitem[{{LSST Science Collaboration} {et~al.}(2009){LSST Science
  Collaboration}, {Abell}, {Allison}, {Anderson}, {Andrew}, {Angel}, {Armus},
  {Arnett}, {Asztalos}, {Axelrod}, \& et~al.}]{LSST}
{LSST Science Collaboration}, {et~al.} 2009, ArXiv e-prints, 0912.0201

\bibitem[{{Maiolino} {et~al.}(2008){Maiolino}, {Nagao}, {Grazian}, {Cocchia},
  {Marconi}, {Mannucci}, {Cimatti}, {Pipino}, {Ballero}, {Calura}, {Chiappini},
  {Fontana}, {Granato}, {Matteucci}, {Pastorini}, {Pentericci}, {Risaliti},
  {Salvati}, \& {Silva}}]{Maiolino08}
{Maiolino}, R., {et~al.} 2008, \aap, 488, 463

\bibitem[{{Mannucci} {et~al.}(2010){Mannucci}, {Cresci}, {Maiolino}, {Marconi},
  \& {Gnerucci}}]{Mannucci10}
{Mannucci}, F., {Cresci}, G., {Maiolino}, R., {Marconi}, A., \& {Gnerucci}, A.
  2010, \mnras, 408, 2115

\bibitem[{{Mannucci} {et~al.}(2011){Mannucci}, {Salvaterra}, \&
  {Campisi}}]{Mannucci11}
{Mannucci}, F., {Salvaterra}, R., \& {Campisi}, M.~A. 2011, \mnras, 439

\bibitem[{{McGaugh}(1991)}]{M91}
{McGaugh}, S.~S. 1991, \apj, 380, 140

\bibitem[{{Modjaz} {et~al.}(2008){Modjaz}, {Kewley}, {Kirshner}, {Stanek},
  {Challis}, {Garnavich}, {Greene}, {Kelly}, \& {Prieto}}]{Modjaz08}
{Modjaz}, M., {et~al.} 2008, \aj, 135, 1136

\bibitem[{{Osterbrock} \& {Ferland}(2006)}]{oandf}
{Osterbrock}, D.~E., \& {Ferland}, G.~J. 2006, {Astrophysics of gaseous nebulae
  and active galactic nuclei}, ed. {Osterbrock, D.~E.~\& Ferland, G.~J.}

\bibitem[{{Pacifici} {et~al.}(2012){Pacifici}, {Charlot}, {Blaizot}, \&
  {Brinchmann}}]{Pacifici12}
{Pacifici}, C., {Charlot}, S., {Blaizot}, J., \& {Brinchmann}, J. 2012, \mnras,
  421, 2002

\bibitem[{{Padmanabhan} {et~al.}(2008){Padmanabhan}, {Schlegel}, {Finkbeiner},
  {Barentine}, {Blanton}, {Brewington}, {Gunn}, {Harvanek}, {Hogg},
  {Ivezi{\'c}}, {Johnston}, {Kent}, {Kleinman}, {Knapp}, {Krzesinski}, {Long},
  {Neilsen}, {Nitta}, {Loomis}, {Lupton}, {Roweis}, {Snedden}, {Strauss}, \&
  {Tucker}}]{Padmanabhan08}
{Padmanabhan}, N., {et~al.} 2008, \apj, 674, 1217

\bibitem[{{Peeples} \& {Shankar}(2011)}]{Peeples11}
{Peeples}, M.~S., \& {Shankar}, F. 2011, \mnras, 417, 2962

\bibitem[{{Perez-Montero} {et~al.}(2012){Perez-Montero}, {Contini},
  {Lamareille}, {Maier}, {Carollo}, {Kneib}, {Le Fevre}, {Lilly}, {Mainieiri},
  {Renzini}, {Scodeggio}, {Zamorani}, {Bardelli}, {Bolzonella}, {Bongiorno},
  {Caputi}, {Cucciati}, {de la Torre}, {de Ravel}, {Franzetti}, {Garilli},
  {Iovino}, {Kampczyk}, {Knobel}, {Kovac}, {Le Borgne}, {Le Brun}, {Mignoli},
  {Pello}, {Peng}, {Presotto}, {Ricciardelli}, {Silverman}, {Tanaka}, {Tasca},
  {Tresse}, {Vergani}, \& {Zucca}}]{PerezM12}
{Perez-Montero}, E., {et~al.} 2012, ArXiv e-prints, 1210.0334

\bibitem[{{Pettini} \& {Pagel}(2004)}]{PP04}
{Pettini}, M., \& {Pagel}, B.~E.~J. 2004, \mnras, 348, L59

\bibitem[{{Pilyugin} {et~al.}(2010){Pilyugin}, {V{\'{\i}}lchez}, \&
  {Thuan}}]{PVT}
{Pilyugin}, L.~S., {V{\'{\i}}lchez}, J.~M., \& {Thuan}, T.~X. 2010, \apj, 720,
  1738

\bibitem[{{Prantzos} \& {Boissier}(2003)}]{PB03}
{Prantzos}, N., \& {Boissier}, S. 2003, \aap, 406, 259

\bibitem[{{Prieto} {et~al.}(2008){Prieto}, {Stanek}, \& {Beacom}}]{Prieto08}
{Prieto}, J.~L., {Stanek}, K.~Z., \& {Beacom}, J.~F. 2008, \apj, 673, 999

\bibitem[{{Rosolowsky} \& {Simon}(2008)}]{Rosolowsky08}
{Rosolowsky}, E., \& {Simon}, J.~D. 2008, \apj, 675, 1213

\bibitem[{{Saglia} {et~al.}(2012){Saglia}, {Tonry}, {Bender}, {Greisel},
  {Seitz}, {Senger}, {Snigula}, {Phleps}, {Wilman}, {Bailer-Jones}, {Klement},
  {Rix}, {Smith}, {Green}, {Burgett}, {Chambers}, {Heasley}, {Kaiser},
  {Magnier}, {Morgan}, {Price}, {Stubbs}, \& {Wainscoat}}]{Saglia12}
{Saglia}, R.~P., {et~al.} 2012, \apj, 746, 128

\bibitem[{{Salim} {et~al.}(2007){Salim}, {Rich}, {Charlot}, {Brinchmann},
  {Johnson}, {Schiminovich}, {Seibert}, {Mallery}, {Heckman}, {Forster},
  {Friedman}, {Martin}, {Morrissey}, {Neff}, {Small}, {Wyder}, {Bianchi},
  {Donas}, {Lee}, {Madore}, {Milliard}, {Szalay}, {Welsh}, \& {Yi}}]{Salim07}
{Salim}, S., {et~al.} 2007, \apjs, 173, 267

\bibitem[{{Sanders} {et~al.}(2012{\natexlab{a}}){Sanders}, {Caldwell},
  {McDowell}, \& {Harding}}]{SandersM31}
{Sanders}, N.~E., {Caldwell}, N., {McDowell}, J., \& {Harding}, P.
  2012{\natexlab{a}}, \apj, 758, 133

\bibitem[{{Sanders} {et~al.}(2012{\natexlab{b}}){Sanders}, {Soderberg},
  {Levesque}, {Foley}, {Chornock}, {Milisavljevic}, {Margutti}, {Berger},
  {Drout}, {Czekala}, \& {Dittmann}}]{nesBS}
{Sanders}, N.~E., {et~al.} 2012{\natexlab{b}}, \apj, 758, 132

\bibitem[{{Sanders} {et~al.}(2012{\natexlab{c}}){Sanders}, {Soderberg},
  {Valenti}, {Foley}, {Chornock}, {Chomiuk}, {Berger}, {Smartt}, {Hurley},
  {Barthelmy}, {Levesque}, {Narayan}, {Botticella}, {Briggs}, {Connaughton},
  {Terada}, {Gehrels}, {Golenetskii}, {Mazets}, {Cline}, {von Kienlin},
  {Boynton}, {Chambers}, {Grav}, {Heasley}, {Hodapp}, {Jedicke}, {Kaiser},
  {Kirshner}, {Kudritzki}, {Luppino}, {Lupton}, {Magnier}, {Monet}, {Morgan},
  {Onaka}, {Price}, {Stubbs}, {Tonry}, {Wainscoat}, \& {Waterson}}]{nes2010ay}
------. 2012{\natexlab{c}}, \apj, 756, 184

\bibitem[{{Schlegel} {et~al.}(1998){Schlegel}, {Finkbeiner}, \& {Davis}}]{SFD}
{Schlegel}, D.~J., {Finkbeiner}, D.~P., \& {Davis}, M. 1998, \apj, 500, 525

\bibitem[{{Stoughton} {et~al.}(2002){Stoughton}, {Lupton}, {Bernardi},
  {Blanton}, {Burles}, {Castander}, {Connolly}, {Eisenstein}, {Frieman},
  {Hennessy}, {Hindsley}, {Ivezi{\'c}}, {Kent}, {Kunszt}, {Lee}, {Meiksin},
  {Munn}, {Newberg}, {Nichol}, {Nicinski}, {Pier}, {Richards}, {Richmond},
  {Schlegel}, {Smith}, {Strauss}, {SubbaRao}, {Szalay}, {Thakar}, {Tucker},
  {Vanden Berk}, {Yanny}, {Adelman}, {Anderson}, {Anderson}, {Annis},
  {Bahcall}, {Bakken}, {Bartelmann}, {Bastian}, {Bauer}, {Berman},
  {B{\"o}hringer}, {Boroski}, {Bracker}, {Briegel}, {Briggs}, {Brinkmann},
  {Brunner}, {Carey}, {Carr}, {Chen}, {Christian}, {Colestock}, {Crocker},
  {Csabai}, {Czarapata}, {Dalcanton}, {Davidsen}, {Davis}, {Dehnen},
  {Dodelson}, {Doi}, {Dombeck}, {Donahue}, {Ellman}, {Elms}, {Evans}, {Eyer},
  {Fan}, {Federwitz}, {Friedman}, {Fukugita}, {Gal}, {Gillespie}, {Glazebrook},
  {Gray}, {Grebel}, {Greenawalt}, {Greene}, {Gunn}, {de Haas}, {Haiman},
  {Haldeman}, {Hall}, {Hamabe}, {Hansen}, {Harris}, {Harris}, {Harvanek},
  {Hawley}, {Hayes}, {Heckman}, {Helmi}, {Henden}, {Hogan}, {Hogg}, {Holmgren},
  {Holtzman}, {Huang}, {Hull}, {Ichikawa}, {Ichikawa}, {Johnston}, {Kauffmann},
  {Kim}, {Kimball}, {Kinney}, {Klaene}, {Kleinman}, {Klypin}, {Knapp},
  {Korienek}, {Krolik}, {Kron}, {Krzesi{\'n}ski}, {Lamb}, {Leger},
  {Limmongkol}, {Lindenmeyer}, {Long}, {Loomis}, {Loveday}, {MacKinnon},
  {Mannery}, {Mantsch}, {Margon}, {McGehee}, {McKay}, {McLean}, {Menou},
  {Merelli}, {Mo}, {Monet}, {Nakamura}, {Narayanan}, {Nash}, {Neilsen},
  {Newman}, {Nitta}, {Odenkirchen}, {Okada}, {Okamura}, {Ostriker}, {Owen},
  {Pauls}, {Peoples}, {Peterson}, {Petravick}, {Pope}, {Pordes}, {Postman},
  {Prosapio}, {Quinn}, {Rechenmacher}, {Rivetta}, {Rix}, {Rockosi}, {Rosner},
  {Ruthmansdorfer}, {Sandford}, {Schneider}, {Scranton}, {Sekiguchi}, {Sergey},
  {Sheth}, {Shimasaku}, {Smee}, {Snedden}, {Stebbins}, {Stubbs}, {Szapudi},
  {Szkody}, {Szokoly}, {Tabachnik}, {Tsvetanov}, {Uomoto}, {Vogeley}, {Voges},
  {Waddell}, {Walterbos}, {Wang}, {Watanabe}, {Weinberg}, {White}, {White},
  {Wilhite}, {Wolfe}, {Yasuda}, {York}, {Zehavi}, \& {Zheng}}]{sloanEDR}
{Stoughton}, C., {et~al.} 2002, \aj, 123, 485

\bibitem[{{Tremonti} {et~al.}(2004){Tremonti}, {Heckman}, {Kauffmann},
  {Brinchmann}, {Charlot}, {White}, {Seibert}, {Peng}, {Schlegel}, {Uomoto},
  {Fukugita}, \& {Brinkmann}}]{Tremonti04}
{Tremonti}, C.~A., {et~al.} 2004, \apj, 613, 898

\bibitem[{{Yates} {et~al.}(2012){Yates}, {Kauffmann}, \& {Guo}}]{Yates12}
{Yates}, R.~M., {Kauffmann}, G., \& {Guo}, Q. 2012, \mnras, 422, 215

\end{thebibliography}

\clearpage

\end{document}